\documentclass[11pt,letterpaper]{llncs}
\usepackage{amsmath}
\usepackage{amssymb}
\usepackage{graphicx}
\usepackage{marvosym}
\usepackage{url}
\usepackage{fancyhdr}

\usepackage{geometry}
\fancypagestyle{firstpage}{\fancyhf{}
\setcounter{page}{1}
\rfoot{\thepage}
}

\usepackage{blindtext}
\usepackage{graphicx}
\usepackage{acronym}
\usepackage[
	n,
	operators,
	advantage,
	sets,
	adversary,
	landau,
	probability,
	notions,	
	logic,
	ff,
	mm,
	primitives,
	events,
	complexity,
	asymptotics,
	keys]{cryptocode}
\usepackage{tikz,pgfplots}
\usepackage{graphicx}
\usepackage{xcolor}
\usepackage{subcaption}
\usepackage{url}
\usepackage{booktabs, multicol, multirow} 
\usepackage{rotating}

\usepackage{enumitem}

\definecolor{headrBlue}{HTML}{92c5de}

\DeclareCaptionType{protocol}[Protocol][]
\newenvironment{prot}{}{}

\newacro{PKC}{Public-key Cryptography}
\newacro{PUF}{Physical Unclonable Function}
\newacro{IoT}{Internet of Things}
\newacro{TTP}{Trusted Third Party}
\newacro{NVM}{Non-volatile Memory}
\newacro{ECC}{Elliptic Curve Cryptography}
\newacro{SRAM}{Static Random-Access Memory}
\newacro{AC}{Activation Code}
\newacro{HD}{Helper Data}
\newacro{IFC}{Integer Factorization Cryptography}
\newacro{FFC}{Finite Field Cryptography}
\newacro{ALU}{Arithmetic Logic Unit}
\newacro{CR}{Challenge-Response}
\newacro{ECDH}{Elliptic Curve Diffie-Hellman}
\newacro{RTL}{Register Transfer Level}
\newacro{IC}{Integrated Circuit}
\newacro{MITM}{Man-In-The-Middle}
\newacro{DH}{Diffie-Hellman}
\newacro{DHKE}{Diffie-Hellman Key Exchange}
\newacro{CMOS}{Complementary Metal Oxide Semiconductor}
\newacro{ECDLP}{Elliptic Curve Discrete Logarithm Problem}
\newacro{DLP}{Discrete Log Problem}
\newacro{GDLP}{Generalized Discrete Logarithm Problem}
\newacro{ECDLP}{Elliptic Curve Discrete Logarithm Problem}
\newacro{SoC}{System on Chip}
\newacro{FPGA}{Field Programmable Gate Array}
\newacro{AXI}{Advanced eXtensible Interface}
\newacro{ASIC}{Application Specific Integrated Circuit}
\newacro{PS}{Processing System}
\newacro{PL}{Programmable Logic}
\newacro{PCB}{Printed Circuit Board}
\newacro{IETF}{Internet Engineering Task Force}
\newacro{RFC}{Request for Comments}
\newacro{AMBA}{Advanced Microcontroller Bus Architecture}
\newacro{PKI}{Public Key Infrastructure}
\newacro{ASIP}{Application-specific Instruction Set Processor}
\newacro{ISA}{Instruction Set Architecture}
\newacro{RNG}{Random Number Generator}
\newacro{GPP}{General Purpose Processor}
\newacro{AES}{Advanced Encryption Standard}
\newacro{ECDSA}{Elliptic Curve Digital Signature Algorithm}
\newacro{APSoC}{All Programmable System on Chip}
\newacro{ROM}{Read-Only Memory}
\newacro{PRNG}{Pseudo Random Number Generator}
\newacro{HDL}{Hardware Description Language}

\begin{document}


\title{\LARGE{Public-Key Based Authentication Architecture for IoT Devices Using PUF}}


%
%


\author{Haji Akhundov\inst{1} \and Erik van der Sluis\inst{2} \and Said Hamdioui\inst{1} \and Mottaqiallah Taouil\inst{1}}

\institute{Delft University of Technology, Delft, The Netherlands\\
\email{H.Akhundov@tudelft.nl}, \email{S.Hamdioui@tudelft.nl}, \email{M.Taouil@tudelft.nl},\\
\and
Intrinsic ID B.V.,
Eindhoven, The Netherlands\\
\email{Erik.van.der.Sluis@intrinsic-id.com}
}




%
%


\maketitle

\thispagestyle{firstpage}

\begin{abstract}
Nowadays, \ac{IoT} is a trending topic in the computing world.
Notably, IoT devices have strict design requirements and are often referred to as \textit{constrained devices}.
Therefore, security techniques and primitives that are lightweight are more suitable for such devices, e.g., \ac{SRAM} \acp{PUF} and \ac{ECC}.
\ac{SRAM} \ac{PUF} is an intrinsic security primitive that is seeing widespread adoption in the IoT segment.
\ac{ECC} is a public-key algorithm technique that has been gaining popularity among constrained IoT devices.
The popularity is due to using significantly smaller operands when compared to other public-key techniques such as RSA (Rivest Shamir Adleman).
This paper shows the design, development, and evaluation of an application-specific secure communication architecture based on \ac{SRAM} \ac{PUF} technology and \ac{ECC} for constrained IoT devices.
More specifically, it introduces an \ac{ECDH} public-key based cryptographic protocol that utilizes PUF-derived keys as the root-of-trust for silicon authentication.
Also, it proposes a design of a modular hardware architecture that supports the protocol.
Finally, to analyze the practicality as well as the feasibility of the proposed protocol, we demonstrate the solution by prototyping and verifying a protocol variant on the commercial Xilinx Zynq-7000 APSoC device.
\end{abstract}

\section{Introduction}


Secure communication has been paramount throughout history \cite{Singh:1999:CBE:519693}.
Although in the early ages it was mainly found in niche applications such as the military and royal society, today it is an inevitable part of our daily lives. 
The recent rapid proliferation of \ac{IoT}, a diverse set of devices that are connected to the Internet, imposes new challenges
for the designers to keep protecting our privacy, security, and safety \cite{IoTproblem}.
Today, the design of use-case specific solutions and its time-to-market are the biggest challenges in this competitive and rapidly developing IoT semiconductor industry, where developers rely on ad-hoc security solutions. 
%
%
%
%
%
%
%
Therefore, there is an urgent need to create cost-effective secure solutions with a short-time development, which are an important facilitator in the IoT market.


Although there is a lot of work published to address the above issues, it mainly focuses on individual aspects such as protocol design or implementations.
One of the first works in this field was in 2004 by Lee et al. \cite{Lee2004} who showed that individual \acfp{IC} can be identified and authenticated using \acfp{PUF}. 
Later, publications such as \cite{Suh2007} described a low-cost authentication protocol of individual ICs using PUF. However, that protocol is basic and is not suitable with so-called weak PUFs \cite{Herder2014}.
More authentication protocols came after that such as in \cite{Frikken2009}, \cite{rfe1}, \cite{Maes2013} and \cite{AGMSY15} and a more recent in \cite{Barbareschi2018}.
Only a few publications that combine both authentication protocols using PUFs and lightweight cost-effective implementations exist in the literature such as \cite{Aysu2016}; the authors developed a lightweight \ac{ASIP} that supports certain existing authentication protocols
based on reverse fuzzy extractor (RFE) constructions, rather than reusing existing components. 
Still, most works typically address the aspects of the authentication protocol
but do not address cost-effectiveness and/or time to market aspects.
A solution that satisfies all the requirements is still needed.



In this work, we focus on developing a cryptographic protocol based on \acf{ECDH} that enables efficient hardware design by reusing readily-available components for an efficient and fast time-to-market design.
In that regard, we designed and developed an efficient and cost-effective solution. 
In short, the contributions of this paper are:
\begin{itemize}
    \item A \textit{protocol} that enables secure communication between constrained devices and a resource-rich party in \textit{untrusted} fields, and using \ac{PUF}-derived keys as the root-of-trust.
    The protocol is based on a conventional \ac{ECDH} key agreement scheme and a fuzzy extractor using code-offset method for accommodating a PUF.
    
    \item A modular \textit{hardware architecture} where the key components are implemented in hardware while minimizing the impact on the silicon footprint.
    Here we emphasize on the fact that readily-available off-the-shelf components such as the NaCl core \cite{Hutter2015} can be used to speed up the development cycle.
    The core can be used to perform elliptic curve scalar multiplications on a patent-free elliptic curve - Curve25519 \cite{Bernstein2006}, offering 128 bits of security.
    
    \item A proof of concept based on a Zynq board to demonstrate how such a solution can be quickly prototyped using `off-the-shelf' components.
\end{itemize}



The rest of this paper is organized as follows.
Section~\ref{sec:backgroung} provides background information.
Section~\ref{sec:protocol} discusses protocol design for our intended application; 
we propose a total of four variants of this protocol, each with its own pros and cons.
Section~\ref{sec:architecture} presents a modular hardware architecture design that supports the protocol.
Section~\ref{seC:poc} gives the proof of concept used for validation.
Finally, Section~\ref{sec:conclusion} provides the conclusion.
\section{Background}
\label{sec:backgroung}

\begin{figure} [t]
    \centering
    \includegraphics[width=0.5\linewidth]{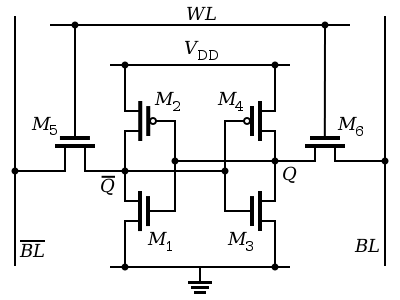}
    \caption{Conceptual schematic of a 6T SRAM cell}
    \label{fig:sram}
\end{figure}
In this section, we provide a  brief background on the working principle of an SRAM-PUF that we use in our work for silicon authentication.

The concept of \ac{PUF}s was first introduced by Pappu \cite{Ravikanth:2001:POF:935173} in 2001 as a \textit{hardware security primitive} that can be used for silicon authentication.
Later, Maes \cite{DBLP:books/sp/Maes13} extended this concept to `expression of an inherent and unclonable instance-specific feature of a physical object.'
In essence, PUFs are functions that take \textit{challenges} as an input and generate \textit{responses} that are random but unique for a specific device \cite{maes2010physically}.
Using PUFs, it is possible to create a stable, unique, and device-dependent fingerprint, which can be used as a secret key or a unique device identifier \cite{cryptoeprint:2015:583}, \cite{6513684}.
Therefore, PUFs are applied in several applications such as anti-counterfeiting, device authentication, and hardware/software binding applications \cite{6513684}.

There is a broad taxonomy of PUFs today, such as delay-based arbiter PUFs and ring oscillator PUFs, memory-based \ac{SRAM} PUFs, and butterfly PUFs \cite{Maes2010}.
SRAM PUFs are of particular interest to the industry compared to other types of PUFs; Integration of SRAMs in the modern systems do not require special manufacturing techniques since SRAMs can be synthesized using standard cells. Furthermore, they are readily available in most existing systems.
\ac{SRAM} \ac{PUF} is a memory-based PUF construction that uses intrinsic random start-up cell values to create challenge pair responses \cite{modelmafalda}.
SRAM is a standard cell component that is composed of 6 transistors.
%
%
%
Figure~\ref{fig:sram} shows a typical six transistor SRAM cell design, consisting of two cross-coupled \ac{CMOS} inverters using four transistors $M_{1}$ through $M_{4}$.
Transistors $M_{5}$ and $M_{6}$ are known as the pass transistors.
The wordline (WL), bitline (BL), and its complement are used to access the cell.
For performance reasons, the two inverters in the SRAM cell are designed in a well-balanced, symmetrical way.
However, the small and random sub-micron process variations in the manufacturing process cause different physical properties of the transistors.
These differences in the transistors of the SRAM cell causes a skew.
Due to this skew, a cell acquires a preferred state of a logic `0' or a logic `1' when powered on, referred to as one bit of `electronic' fingerprint.
This phenomenon of inherent, device-unique variations makes SRAM PUFs construction possible \cite{Maes2010}.
It is resistant to cloning even if one can get their hands on the circuit design/layout files since the skew is not visible in the layout.
With the current manufacturing process variations are inevitable and cannot be controlled; therefore, cloning an SRAM PUF yields to be tough or even impossible \cite{DBLP:books/sp/Maes13}.
Every SRAM cell upon power-up can provide one bit of such electronic fingerprint.
%
%
Hence, arrays of \textit{uninitialized} SRAM cells can be used to identify devices, securely store, and generate cryptographic keys on devices.

In this work, we use SRAM \ac{PUF}-derived secret key as the hardware root-of-trust in authenticating constrained IoT devices in the field. 
Note that each time the key must be reliably extracted.
Techniques such as \textit{entropy extraction} and \textit{error-correcting codes} are used to achieve this \cite{cryptoeprint:2015:583}.
\section{Protocol Design}
\label{sec:protocol}

In this section, we show how a secure protocol can be efficiently composed for an IoT application. 
First, we describe what confidentiality and authentication are, and then provide a realistic use case. After that, we present the main protocol and variants thereof. Finally, we discuss the advantages and disadvantages of the proposed protocols.

\textit{Confidentiality and authentication} are some of the core criteria of a secure system \cite{kumar2006elliptic}.
\textit{Confidentiality} is a service used to keep the information accessible only to authorized users.
\textit{Authentication} is a service that verifies the identity of users or entities and therefore ensures that its data or the entity can be trusted.
In order to achieve confidentiality and authentication between devices, one may employ authenticated encryption techniques.
However, before establishing an encrypted channel, communicating parties must share the same key.
Due to the key distribution problem \cite{Paar:2009:UCT:1721909}, key exchange protocols have emerged.
Using a key exchange (key-agreement) protocol, involved parties agree on a shared secret key in such a way that all parties influence the outcome, without transferring the actual key itself over an untrusted channel.
Key-agreement protocols rely on the exchange of authentic public-key components of the involved parties, i.e., keys that truly belong, therefore prove the identity of claimed users or entities.
Next, we present a use case where a key-agreement protocol is used to achieve this.

\subsection{Use Case}

\setlength{\belowcaptionskip}{-10pt}
\begin{figure}
    \centering
    \includegraphics[scale=0.9]{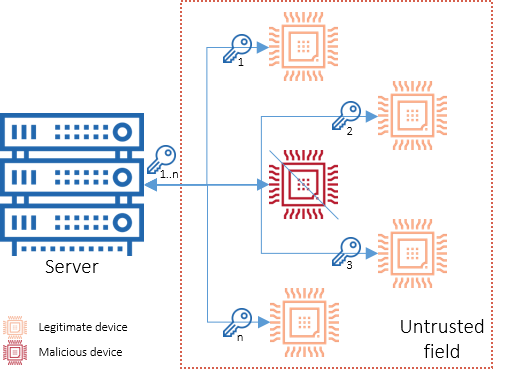}
    \caption{Use-case scenario}
    \label{fig:use_case_application}
\end{figure}

The use case under consideration in this paper is illustrated in Figure~\ref{fig:use_case_application};  a resource-rich server communicates with constrained IoT devices in an untrusted field.
The two fundamental assumptions about this untrusted field are that the communication is susceptible to both \textit{passive and active attacks} because there is very little or no control over it.
In a passive attack, an intruder can only eavesdrop on the communication.
In an active attack, an intruder may also transmit, replay, modify or delete specific communication messages.
We see that there are two types of devices: legitimate and malicious.
The latter should not be authenticated.
The need for \textit{secure communication} in this use case is, therefore, obvious.

\subsection{Protocol Description}

As pointed out in the background section, establishing secure communication between a device and a server requires a secure protocol.
The goal of the protocol is to derive shared keys between the server and legitimate devices in the field, therefore, enabling an authenticated and encrypted communication.
We will briefly look at four slightly different scenarios with different requirements that result in protocol modifications and discuss their advantages/disadvantages.
In all the scenarios, the communication is to be established between a server and a device enrollment.
By using a \ac{TTP}, we add substantial flexibility due to the \ac{PKI} and therefore inherit the benefit of certificate management, e.g., device's certificate revocation, etc.
The four scenarios are:
\begin{itemize}[leftmargin=*]
\item \textit{Scenario 1:} 
we wish to achieve mutual authentication and certificate management, and use as little \acf{NVM} as possible on the device.
To achieve mutual authentication, the server and devices need to be enrolled by a \ac{TTP}.
To reduce \ac{NVM} requirement and improve certificate management, cloud infrastructure is used.
The cloud infrastructure is a database that is meant to be accessible by the \ac{TTP} and the server for storing and retrieving digital certificates of the enrolled devices, respectively.
\item \textit{Scenario 2:} 
Therefore, similarly as in scenario 1, both parties are enrolled by a \ac{TTP}; however, cloud infrastructure is not used here.
\item \textit{Scenario 3:} 
we wish to achieve one-way authentication (which could be sufficient in some instances) and certificate management, and again use as little as possible of device's \ac{NVM}.
To achieve this, only the devices are enrolled by the \ac{TTP}, and cloud infrastructure is used.
\item \textit{Scenario 4:} 
we loosen up all the constraints, i.e., one-way authentication, no reduced \ac{NVM} requirement, and no certificate management is required.
Clearly, this is the minimal version of the protocol, requiring only the enrollment of the devices and no cloud infrastructure.
\end{itemize}

For each of the scenarios, we present four different protocol variants, referred to as Protocol A, B, C, and D, respectively.
In the rest of the section, we focus on protocol Variant A that satisfies the most demanding Scenario 1 in detail, shown in Protocol~\ref{prot:variant_a}.
Other protocol variants are briefly explained afterward.

\begin{figure*}[ht]
    \begin{center}
    \begin{prot}
        \resizebox{1\textwidth}{!} {
        \fbox{
        \pseudocode[codesize=\large]
        {%
        \textbf{(I) Enrollment of the device and the server} \< \< \\[0.2\baselineskip]
        \textbf{Trusted Third Party (TTP)}~has~SK_{TTP} \<\< \textbf{Device~(ID)} \\[0.1\baselineskip][\hline]
         \<\<\<\\ [-0.5\baselineskip]
         \<\<\pcln (x,~HD)~\gets PUF-enroll\\
         \<\<\pcln PK_{ID}\gets ~DH~\kgen(x) \\ 
         \< \sendmessageleft * [6 cm] {\pcln ID, HD, PK_{ID} } \<\\
         \pcln \sigma_{ID} \gets \sig_{\sk_{TTP}}(ID, HD, PK_{ID}) \<\<\\
         \pcln Cert_{ID}=[ID, HD, PK_{ID}, \sigma_{ID}] \<\<\\
        \pcln Store(Cert_{ID})~in~cloud \<\<\\
        \< \sendmessageright * [6 cm]{\pcln PK_{TTP}} \<\\
         \<\< \pcln Store(PK_{TTP})\\
        \textbf{Trusted Third Party (TTP)}~has~SK_{TTP} \<\< \textbf{Server~(ID)} \\[0.1\baselineskip][\hline]
         \<\<\<\\ [-0.5\baselineskip]
         \<\<\pcln (SK_{server},PK_{server})\gets~DH~ \kgen \\
         \< \sendmessageleft * [6 cm] {\pcln ID, PK_{server} } \<\\
         \pcln \sigma_{ID} \gets \sig_{\sk_{TTP}}(ID, PK_{server}) \<\<\\
         \pcln Cert_{ID_{server}}=[ID, PK_{ID}, \sigma_{ID}] \<\<\\
         \< \sendmessageright * [6 cm]{\pcln Cert_{ID_{server}},PK_{TTP}} \<\\
         \<\< \pcln Store(Cert_{ID_{server}},SK_{server}, PK_{TTP})\\ 
         \textbf{(II) Key-agreement and Authentication}  \<\< \\[0.2\baselineskip]
         \textbf{Server}~has~PK_{TTP}~and~access~to~cloud \<\< \textbf{Device~(ID)}~has~PK_{TTP}  \\[0.1\baselineskip][\hline]
         \<\<\<\\ [-0.5\baselineskip] 
         \pcln Retrieve(Cert_{ID})~from~cloud\\
         \pcln \verify_{\pk_{TTP}}(Cert_{ID}) \\
         \< \sendmessageright * [6 cm]{\pcln Session~request,~HD,~Cert_{ID_{server}}} \<\\
         \<\<\pcln \verify_{\pk_{TTP}}(Cert_{ID_{server}})\\
                \<\< \pcln x~\gets PUF-reconstruct(HD)\\ 
         \pcln w=SK_{server}*PK_{ID} \<\<  \pcln w=x*PK_{server}\\
         \pcln \key~=KDF(w) \<\< \key~=KDF(w)  \\
         \pcln \text{Challenge-response handshake in both directions} \<\< 
        }
     }
    }

    \end{prot}
%
\end{center}
\captionof{protocol}{Protocol based on \acf{DHKE} using \acf{PUF}-derived key. Variant A - \textit{Achieving mutual-authentication and low \acf{NVM} requirement using the cloud infrastructure.}}
\label{prot:variant_a}
\end{figure*}

\noindent\textbf{\textit{Variant A:}}
The protocol is divided into two stages.
In the first stage, the device is enrolled by a \acf{TTP} in a secure environment.
This is typically done only once during the life-cycle of a device.
The second stage, key-agreement, and authentication, takes place in the field whenever necessary, i.e., communicating parties establish a shared key before communication.
Table~\ref{tab:legend} provides the legend for the protocols used in this paper.
Furthermore, other notations, followed by a detailed description of Protocol~\ref{prot:variant_a} are described below:

\begin{table}[h!]
\caption{Protocol Legend}
\label{tab:legend}
    \begin{center}
        \begin{tabular}{|c|c|c|c|}
            \hline  
              \textbf{Symbol} &  \textbf{Description} &
              \textbf{Symbol} &  \textbf{Description} \\\hline
            $ID$ &  Device ID, 48 bits &
            $SK_{TTP}$ &  TTP's secret key, 256 bits\\\hline
            
            $PK_{TTP}$ &  TTP's public key, 256 bits &
            $x$ &  PUF-based secret key, 256 bits \\\hline
            $HD$ &  Helper data, 752 bytes &
            $PK_{ID}$ &  Device's Public Key, 256 bits\\\hline
            $\sigma_{ID}$ &  Digital signature, 256 bits &
           
            $SK_{server}$ & Server's secret key, 256 bits\\\hline
            $PK_{server}$ & Server's public key, 256 bits &
            $w, k$ &  Shared secret and key, 256 bits\\\hline
            
            $Cert_{ID}$ &  Device's Certificate  &
            $Cert_{ID_{server}}$ &  Certificate \\\hline
            
            $*$ &  Scalar multiplication &
            $KDF()$ &  Key Derivation Function \\\hline
            $Vf$ &  Signature Verification & & \\\hline
        \end{tabular}
    \end{center}
\end{table}

            
           
            
            

\vspace{1ex}
 \begingroup
 \setlength{\parindent}{0.0em}

\paragraph*{PUF-enroll:} generates a (PUF-derived) cryptographically secure key $x$ and helper data $HD$ that is used in the decoding stage of the key reconstruction in the field.

\paragraph*{PUF-reconstruct:} is the reconstruction process of the secure key $x$ that was enrolled earlier.

\paragraph*{DH KGen($x$):} calculates the public key based on the private key $x$. In \ac{ECC}, it corresponds to a scalar multiplication (*) of the scalar $x$ and the base point of a particular elliptic curve as per curve's specifications.

\paragraph*{KDF():} is a Key Derivation Function that should be used to derive a quality secret key from a shared secret \cite{Hugo2010}.

\paragraph*{ID:} is an identification number unique to a device.

    
    
        

    
    
\endgroup

\textbf{(I) Enrollment}
\begingroup
\setlength{\parindent}{0.0em}
All IoT devices must first be enrolled with a \ac{TTP} before their operation in the field.
The enrollment phase happens as follows:
\paragraph*{Step 1:} Device's PUF response is enrolled, and the key-generation subsystem generates a cryptographic key $x$, along with \ac{HD}.
The HD is required to reconstruct the private key $x$ from the same \ac{PUF} device in later stages.
This is explained in more detail in Section~\ref{sec:high_level}.
\paragraph*{Step 2:} The generated key $x$ is used to calculate its corresponding public key.
This computation is performed using scalar multiplication in a suitable elliptic curve group, the result of which is a point on the curve.
Note that in general, depending on the cryptosystem, $x$ may not be directly used as a private key. A post-processing step may be needed.
\paragraph*{Step 3:} The device sends its identifier \textit{ID}, \textit{\ac{HD}} and the computed public key to the \ac{TTP}.
\paragraph*{Step 4-5:} The \ac{TTP} signs the received data using its private key and generates a certificate.
\paragraph*{Step 6:} TTP stores the device's certificate to a cloud.
The certificate binds the device's ID to its \ac{PUF} based public key.
\paragraph*{Step 7-8:} The device must be able to verify the identity of the server it is trying to communicate within the field by verifying its certificate.
Therefore, an additional step in device enrollment is sending and storing the \ac{TTP}'s public key $PK_{TTP}$ on the device.
To guarantee the security of this protocol, this key must be stored securely on the device, i.e., it cannot be tampered with.
Failing to do so allows malicious servers to masquerade as a trusted one.
\paragraph*{Step 9-14:} In order to achieve mutual authentication, the server needs to be enrolled by a \ac{TTP} similarly as the device; a digital certificate is issued to the server. 

\endgroup
\textbf{(II) Key-agreement and Authentication}
\begingroup
\setlength{\parindent}{0.0em}
In the field, the device must be able to establish a secure and authenticated communication with the server.
\paragraph*{Step 15-16:} The server retrieves the certificate of the desired device from the cloud and verifies it.
\paragraph*{Step 17:} The server initiates communication by sending a \textit{session request} message, HD of the device, and the certificate to the device for verification.
\paragraph*{Step 18:} The device verifies server's certificate using  $PK_{TTP}$.
\paragraph*{Step 19:} The device reconstructs the \ac{PUF}-based private key $x$ using the received HD.
\paragraph*{Step 20-21:} Both parties have the required information to calculate the shared secret $w$ using scalar multiplication.
\paragraph*{Step 22:} A \textit{Key Derivation Function} is used for privacy-enhancing purposes to derive a cryptographically secure shared key on both sides.
%
%
\paragraph*{Step 23:} A challenge-response handshake in both directions is necessary to make sure that the calculated shared keys on both sides are equivalent.

\endgroup
\subsection{Other Variants}

Alternatively, if one wishes to have simpler protocol variants (Scenarios 2-4), i.e., no cloud infrastructure or simply only one-way authentication, one may use a simpler protocol.
In this subsection, we present three variants with different complexities; similarly to Variant A, all of them are divided into two stages.

\begin{figure*}[t]
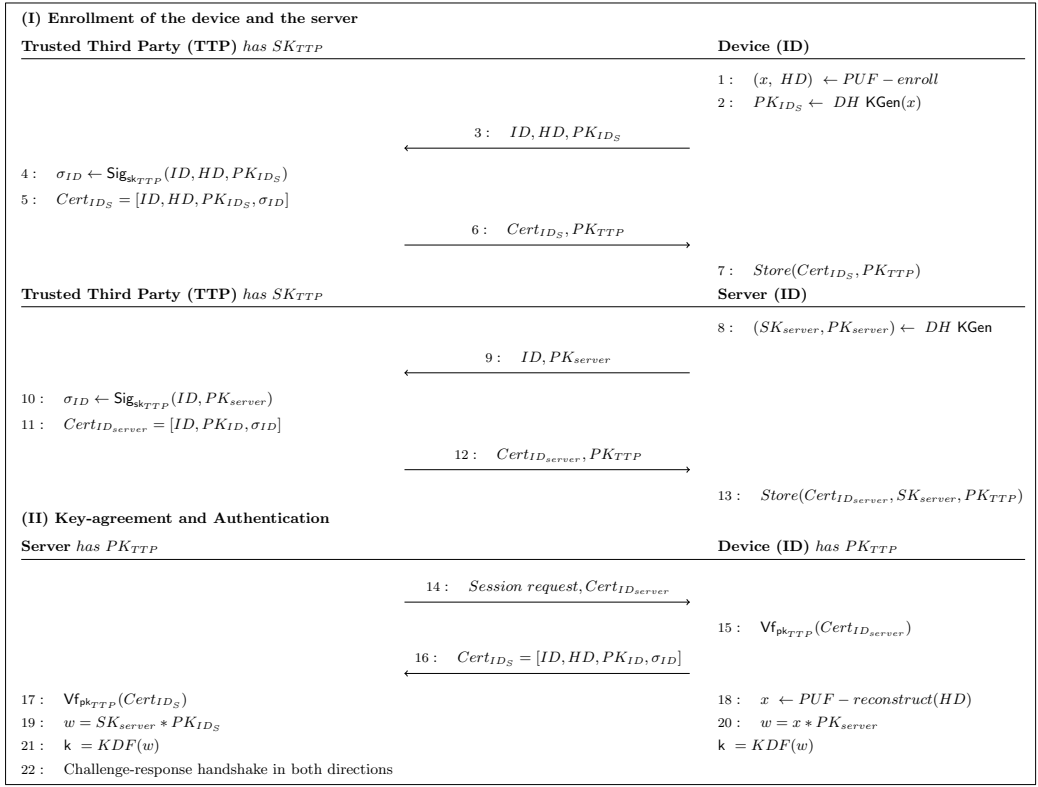

    \begin{center}
    \begin{prot}
        \resizebox{1\textwidth}{!} {
        \fbox{
            \pseudocode{%
            \textbf{(I) Enrollment of the device and the server} \< \< \\[0.2\baselineskip]
            \textbf{Trusted Third Party (TTP)}~has~SK_{TTP} \<\< \textbf{Device~(ID)} \\[0.1\baselineskip][\hline]
             \<\<\<\\ [-0.5\baselineskip]
             \<\<\pcln (x,~HD)~\gets PUF-enroll\\
             \<\<\pcln PK_{ID_S}\gets ~DH~\kgen(x) \\ 
             \< \sendmessageleft * [6 cm] {\pcln ID, HD, PK_{ID_S} } \<\\
             \pcln \sigma_{ID} \gets \sig_{\sk_{TTP}}(ID, HD, PK_{ID_S}) \<\<\\
             \pcln Cert_{ID_{S}}=[ID, HD, PK_{ID_S}, \sigma_{ID}] \<\<\\
             \< \sendmessageright * [6 cm]{\pcln Cert_{ID_{S}},PK_{TTP}} \<\\
             \<\< \pcln Store(Cert_{ID_{S}},PK_{TTP})\\ 
            \textbf{Trusted Third Party (TTP)}~has~SK_{TTP} \<\< \textbf{Server~(ID)} \\[0.1\baselineskip][\hline]
             \<\<\<\\ [-0.5\baselineskip]
             \<\<\pcln (SK_{server},PK_{server})\gets~DH~ \kgen \\
             \< \sendmessageleft * [6 cm] {\pcln ID, PK_{server} } \<\\
             \pcln \sigma_{ID} \gets \sig_{\sk_{TTP}}(ID, PK_{server}) \<\<\\
             \pcln Cert_{ID_{server}}=[ID, PK_{ID}, \sigma_{ID}] \<\<\\
             \< \sendmessageright * [6 cm]{\pcln Cert_{ID_{server}},PK_{TTP}} \<\\
             \<\< \pcln Store(Cert_{ID_{server}},SK_{server}, PK_{TTP})\\ 
             \textbf{(II) Key-agreement and Authentication}  \<\< \\[0.2\baselineskip]
             \textbf{Server}~has~PK_{TTP} \<\< \textbf{Device~(ID)}~has~PK_{TTP}  \\[0.1\baselineskip][\hline]
             \<\<\<\\ [-0.5\baselineskip] 
             \< \sendmessageright * [6 cm]{\pcln Session~request,Cert_{ID_{server}}} \<\\
             \<\<\pcln \verify_{\pk_{TTP}}(Cert_{ID_{server}})\\
             \< \sendmessageleft * [6 cm]{\pcln Cert_{ID_{S}}=[ID, HD, PK_{ID}, \sigma_{ID}]} \<\\
             \pcln \verify_{\pk_{TTP}}(Cert_{ID_S}) \<\< \pcln x~\gets PUF-reconstruct(HD)\\ 
             \pcln w=SK_{server}*PK_{ID_S} \<\<  \pcln w=x*PK_{server}\\
             \pcln \key~=KDF(w) \<\< \key~=KDF(w)  \\
            \pcln \text{Challenge-response handshake in both directions} \<\< 
             }
         }
        }
    \captionof{protocol}{Protocol based on \ac{DHKE} using \ac{PUF}-derived key. Variant B - \textit{Mutual-authentication and no cloud infrastructure.}}
    \label{prot:variant_b}
    \end{prot}
    \end{center}
\end{figure*}

\noindent\textbf{\textit{Variant B:}}
This protocol variant, shown in Protocol~\ref{prot:variant_b}, accommodates Scenario 2; therefore, it does not require a cloud infrastructure, while still having mutual authentication.
The cloud infrastructure is removed in this variant.
The certificate that is generated in Step 5 is transmitted to and stored on the device itself, therefore requiring more NVM storage space.
Instead of Step 15, upon a session request, the device sends its previously-stored certificate to the server.
    
\noindent\textbf{\textit{Variant C:}}
This protocol variant, shown in Protocol~\ref{prot:variant_c}, accommodates Scenario 3, therefore, mutual authentication is removed, i.e., Steps 7 - 14 are omitted.
This protocol uses cloud infrastructure to reduce the NVM storage requirement on the device and provides certificate management.
Step 17 is modified, and Step 18 is removed since no server certificates are involved.
The server fetches and verifies the certificate from the cloud, generates an ephemeral DH key-pair, and sends its contribution ($PK_{server}$) to the device.
At this point, both parties can start the key generation process, as seen in Steps 19-23 of Variant A.
Note that a challenge-response handshake in one direction is sufficient here.
\begin{figure*}[t]
    \begin{center}
    \begin{prot}
        \resizebox{1\textwidth}{!} {
        \fbox{
            \pseudocode{%
            \textbf{(I) Enrollment of the Device} \< \< \\[0.2\baselineskip]
            \textbf{Trusted Third Party (TTP)}~has~SK_{TTP} \<\< \textbf{Device~(ID)} \\[0.1\baselineskip][\hline]
             \<\<\<\\ [-0.5\baselineskip]
             \<\<\pcln (x,~HD)~\gets PUF-enroll\\
             \<\<\pcln PK_{ID}\gets ~DH~\kgen(x) \\ 
             \< \sendmessageleft * [6 cm] {\pcln ID, HD, PK_{ID} } \<\\
             \pcln \sigma_{ID} \gets \sig_{\sk_{TTP}}(ID, HD, PK_{ID}) \<\<\\
             \pcln Cert_{ID}=[ID, HD, PK_{ID}, \sigma_{ID}] \<\<\\
            \pcln Store(Cert_{ID})~in~cloud \<\<\\
             \textbf{(II) Key-agreement and Authentication}  \<\< \\[0.2\baselineskip]
             \textbf{Server}~has~PK_{TTP}~and~access~to~cloud~storage \<\< \textbf{Device~(ID)}  \\[0.1\baselineskip][\hline]
             \<\<\<\\ [-0.5\baselineskip] 
             \pcln Retrieve(Cert_{ID})~from~cloud\\
             \< \sendmessageright * [6 cm]{\pcln Session~request,~HD} \<\\
             \pcln \verify_{\pk_{TTP}}(Cert_{ID}) \<\< \pcln x~\gets PUF-reconstruct(HD)\\ 
             \pcln (SK_{HP},PK_{HP})\gets Ephemeral~DH~ \kgen \<\<\\
             \< \sendmessageright * [6 cm]{\pcln PK_{HP}} \<\\
             \pcln w=SK_{HP}*PK_{ID} \<\<  \pcln w=x*PK_{HP}\\
             \pcln \key~=KDF(w) \<\< \key~=KDF(w)  \\
            \pcln \text{Challenge-response handshake in one directions} \<\< 
             }
         }
        }
    \captionof{protocol}{Protocol based on \ac{DHKE} using \ac{PUF}-derived key. Variant C - \textit{Device authentication and using cloud infrustructure.}}
    \label{prot:variant_c}
    \end{prot}
    \end{center}
\end{figure*}

\noindent\textbf{\textit{Variant D:}}
This protocol variant, shown in Protocol~\ref{prot:variant_d}, accommodates Scenario 4, therefore is the simplest variant of the protocols.
In this variant, only the device is enrolled, and the certificate that is generated in Step 5 is transmitted to and stored on the device itself.
Therefore, Steps 6-16 are omitted.
Steps 17-18 are modified as follows; upon a session request, the device sends the certificate back that was stored on the device during enrollment.
The server verifies the certificate, generates an ephemeral DH key-pair, and sends its contribution ($PK_{server}$) to the device.
At this point, both parties can start the key generation process, as seen in Steps 19-23 of Variant A.
Note that a challenge-response handshake in one direction is sufficient in this case.
\begin{figure*}[t]
    \begin{center}
        \begin{prot}
            \resizebox{1\textwidth}{!} {
                \fbox{
                    \pseudocode{%
                        \textbf{(I) Enrollment of the Device} \< \< \\[0.2\baselineskip]
                        \textbf{Trusted Third Party (TTP)}~has~SK_{TTP} \<\< \textbf{Device~(ID)} \\[0.1\baselineskip][\hline]
                        \<\<\<\\ [-0.5\baselineskip]
                        \<\<\pcln (x,~HD)~\gets PUF-enroll\\
                        \<\<\pcln PK_{ID}\gets ~DH~\kgen(x) \\ 
                        \< \sendmessageleft * [6 cm] {\pcln ID, HD, PK_{ID} } \<\\
                        \pcln \sigma_{ID} \gets \sig_{\sk_{TTP}}(ID, HD, PK_{ID}) \<\<\\
                        \pcln Cert_{ID}=[ID, HD, PK_{ID}, \sigma_{ID}] \<\<\\
                        \< \sendmessageright * [6 cm]{\pcln Cert_{ID}} \<\\
                        \<\< \pcln Store(Cert_{ID})\\ 
                        [0.1\baselineskip] \pclb
                        \pcintertext [dotted] {Device enrollment complete} \\
                        \textbf{(II) Key-agreement and Authentication}  \<\< \\[0.2\baselineskip]
                        \textbf{Server}~has~PK_{TTP} \<\< \textbf{Device~(ID)}  \\[0.1\baselineskip][\hline]
                        \<\<\<\\ [-0.5\baselineskip] 
                        \< \sendmessageright * [6 cm]{\pcln Session~request} \<\\
                        \< \sendmessageleft * [6 cm]{\pcln Cert_{ID}=[ID, HD, PK_{ID}, \sigma_{ID}]} \<\\
                        \pcln \verify_{\pk_{TTP}}(Cert_{ID}) \<\< \pcln x~\gets PUF-reconstruct(HD)\\ 
                        \pcln (SK_{HP},PK_{HP})\gets Ephemeral~DH~ \kgen \<\<\\
                        \< \sendmessageright * [6 cm]{\pcln PK_{HP}} \<\\
                        \pcln w=SK_{HP}*PK_{ID} \<\<  \pcln w=x*PK_{HP}\\
                        \pcln \key~=KDF(w) \<\< \key~=KDF(w)  \\
                        \pcln \text{Challenge-response handshake in one directions} \<\< 
                    }
                }
            }
            \captionof{protocol}{Protocol based on \ac{DHKE} using \ac{PUF}-derived key. Variant D - \textit{Device authentication and no cloud infrustructure.}}
            \label{prot:variant_d}
        \end{prot}
    \end{center}
\end{figure*}


\subsection{Protocol Evaluation}

The fundamental security of the proposed protocols relies on the fact that they are built based on the well known \ac{ECDH} protocol.
That said, several assumptions need to be in place to guarantee security.
Firstly, we assume that \ac{TTP} is indeed trusted and that $SK_{TTP}$ is well protected.
Secondly, in Protocols A and B, the storage of \ac{TTP}'s public key $PK_{TTP}$ on the device must be secure.
Although $PK_{TTP}$ is public information, it must not be tampered; if tampered, the security would be compromised, i.e., a malicious server would be able to communicate with the device.
Lastly, storing the device certificate $Cert_{ID}$ on the device or in the cloud does not need to be that secure, because $Cert_{ID}$ is public information.
Stealing its contents will not give any advantage to anyone.
However, modifying it would render it useless, and could be used to perform a denial of service attack on the device, which is why secure storage is still recommended.

Although the differences between the protocol variants are slight changes, the results in terms of communication performance, functionality, etc. can be significantly different, as we will see.
The main properties of the protocol variants A-D are summarized in Table~\ref{tab:properties}.
Options/properties with (+) are desired, whereas (-) are not.
From the table, we see that in all protocol variants, authenticate the device, whereas only Variants A and B authenticate the server as well. 
The \ac{NVM} requirement is discussed later in more detail.
Another important point to note is that Variants C and D generate ephemeral DH key-pairs, i.e., two same parties will set up new keys for every session, a property that is desired.
Alas, protocol Variants A and B do not possess this quality anymore because the signed long term keys are used in the DH handshake.
A quick fix for that would be to generate an ephemeral DH key-pair for the session and use the long-term signed key to sign the new ephemeral key.
The device will then have to verify this signature.
Table~\ref{tab:basic_measurements} shows some basic properties of the protocol variants such as the number of transfers, data transfer size, and the \ac{NVM} requirement to perform the enrollment, key-agreement, and authentication. Note that the numbers are based on the following realistic considerations. 
The size of ID is chosen to be identical to the size of a Media Access Control address (IEEE Standard), which is 48 bits.
A key with 256 bits of entropy needs approximately 720 bytes of SRAM and 752 bytes of HD based on one of the specific implementations of Intrinsic-ID Quiddikey PUF technology. 
Moreover, the following analysis is focused more on the device side and the interactions with the device due to its constrained nature.

The \textit{Number of Transfers} shows the number of message transactions during enrollment and in the field.
As we can see, Variant B has the most data transfers, whereas Variant C has the least.
However, the difference is mainly in the enrollment phase and not during operation; hence, the performance during run-time is similar between them.

The \textit{Data Transfer Size} shows the size of the messages in bits needed to be communicated during the transactions.
This metric is essential for constrained devices because every bit sent consumes power.
Interestingly, all protocol variants transfer more or less the same amount of data.


The \textit{NVM Requirement} shows how much data need to be `permanently' stored on the device.
Variant B requires the most since it needs to store the certificate as well as the \ac{TTP}'s public key, whereas Variant C requires no storage.

\begin{table} 
  \centering
  \caption{Distinguishing Properties of Protocol Variants A-D}%
  \label{tab:properties}%
   \resizebox{\columnwidth}{!}{%
    \begin{tabular}{l|llllll}
        \toprule
        \multicolumn{1}{c}{Protocol} & \begin{sideways}Device Authentication\end{sideways} & \begin{sideways}Server Authentication\end{sideways} & \begin{sideways}NVM Requirement\end{sideways} & \begin{sideways}Cloud Infrastructure\end{sideways} & \begin{sideways}Certificate Management\end{sideways} & \begin{sideways}Sig. Verification
        on Device\end{sideways} \\
        \midrule
        Variant A & +  & +  & negligible (+)  & required (+/-) & online (+) & required (--)  \\
        Variant B & +  & +  & large (--) &  & offline (+/-) & required (--)\\
        Variant C & +  & -  & none (++) &  required (+/-) & online (+) &  \\
        Variant D & +  & -  & large (--)  &  & offline (+/-) &  \\
        \bottomrule
    \end{tabular}%
    }
\end{table}%

\begin{table}[t]
    \centering
    \caption{Properties of Protocol Variants A-D}
    \label{tab:basic_measurements}%
    \resizebox{\columnwidth}{!}{%
        \begin{tabular}{l|llll}
            \toprule
                        & Variant A & Variant  B & Variant C & Variant D  \\ \hline
            Number of Transfers & & & & \\
            \multicolumn{1}{c|}{Stage I} & 4 & 4 & 1 & 2 \\
            \multicolumn{1}{c|}{Stage II} & 4 & 5 & 4 & 5 \\
            \multicolumn{1}{r|}{Total:} & 8 & 9 & 5 & 7 \\ \hline
            Data Transfer Size in bits (with device only) & & & & \\
            \multicolumn{1}{c|}{Stage I} & 6576 & 6832 & 6320 & 6576 \\
            \multicolumn{1}{c|}{Stage II} & 6736 & 7296 & 6368 & 6928 \\
            \multicolumn{1}{r|}{Total:} & 13312 & 14128 & 12688 & 13504 \\ \hline
            NVM Requirement (device) & 256 & 6832 & 0 & 6576 \\
                        & \{PKttp\} & \{Certid\} \{PKttp\} & & \{Certid\} \\
            \bottomrule
        \end{tabular}%
    }
\end{table}%
\section{Architecture Design}
\label{sec:architecture}
In the previous section, a key-exchange protocol based on \acf{ECDH} and \acf{PUF}-derived key has been proposed along with several protocol variants.
In this section, we design a template of a hardware architecture that enables these protocols on a constrained IoT device. The template contains vital components that are necessary to secure the application. The template enables a quick design of high-level hardware architecture.
After that, we present this high-level system architecture and the components that it is comprised of, followed by a proof-of-concept validation in the next section. 

\label{sec:high_level}

\begin{figure}[t]
    \centering
    \includegraphics[width=0.5\columnwidth]{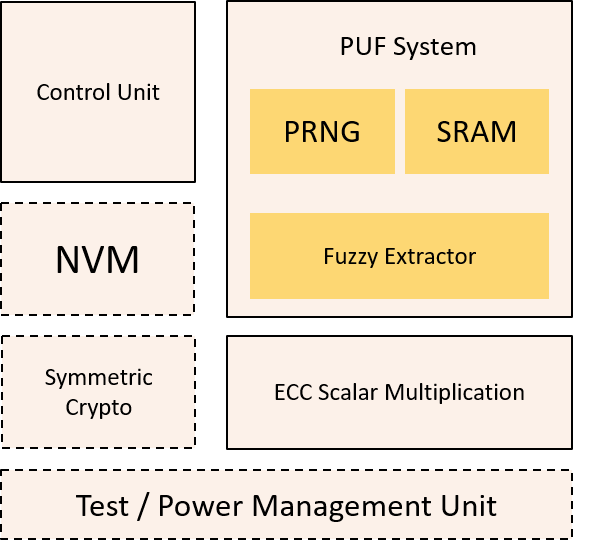}
    \caption{Conceptual Hardware Architecture}
    \label{fig:template}
\end{figure}

The constrained device must include a minimal set of primitives as elaborated below and shown in Figure~\ref{fig:template}:

\noindent \textbf{Control unit:}
    A control unit is essential to orchestrate all components and interface with the outside world.
    For example, the control unit can be a micro-program that implements the protocol and handles the interface to the outside.

\noindent \textbf{ECC Scalar Multiplication unit:}
    The protocols are based on \ac{ECDH}; hence, scalar multiplication operation $*$ is used on the device both during enrollment and in the field.
    The scalar multiplication is the most compute-intensive operation in \ac{ECC}.

\noindent \textbf{PUF System:}
    The protocol is based on \ac{PUF}-derived keys, i.e.,  PUF-technology is used for silicon authentication.
    The selected \ac{PUF} is an \ac{SRAM}-PUF due to its availability in most systems.
    The following are the integral parts of the SRAM-PUF system:
        \begin{itemize}[leftmargin=*]
            \item \textbf{\ac{SRAM}:} 
            For an \ac{SRAM} PUF, a block of uninitialized SRAM must be available in the system.
            \item \textbf{Fuzzy Extractor:}
            SRAM start-up values are typically noisy, mostly due to environmental factors, e.g., temperature variation.
            To compensate for this noise, a fuzzy extractor is used for a reliable and stable secure-key reconstruction \cite{Boyen:2004:RCF:1030083.1030096,Dodis2004,Linnartz2003}.
            A detailed description of a fuzzy extractor, alongside its security or reliability evaluation, is outside the scope of this work. 
            For the sake of proof of concept, we use the simple code offset method, which constitutes of two phases, \textit{enrollment} and \textit{reconstruction}.
            During enrollment, \acf{HD} $HD$ is calculated as $HD=R~\oplus~C   = R~\oplus~Encode(S) $, where $R$ is the initial PUF response and $C$ is the code, which is the result of an encoded secret key $S$.
            During reconstruction, the noisy PUF response $R'$ is read out and the noisy code $C'$ is reconstructed using $ C'=R'~\oplus~HD   = R'~\oplus~(R~\oplus~Encode(S))  =  noise+Encode(S) $
            Eventually, the secret key $S$ is obtained by $ S=Decode(noise+Encode(S)) $
            \item \textbf{\ac{PRNG}:}
            Essentially, SRAM-PUF is used as a key-storage.
            During the PUF enrollment stage, a key $S$ must be supplied to the fuzzy-extractor to be programmed.
            One such source of a key can be a PRNG.
            PRNG needs to be seeded with a TRNG. For that, the noise in the PUF responses could be used.
        \end{itemize}

\setlength{\belowcaptionskip}{-10pt}

The above list represents a minimalist set of components. Optionally, based on the protocol variant the final design might include additional modules such as a \textit{test unit}, \textit{\acf{NVM}}, \textit{symmetric crypto unit}, and a \textit{power management unit}.
\section{Proof of Concept}
\label{seC:poc}
In this section we present the proof of concept that is used for validation as well as a discussion.

In order to demonstrate the practicality of the proposed protocol, we build a prototype using off-the-shelf components that satisfies all requirements.
As a prototyping platform, we choose the Xilinx Zynq-7000 family \ac{APSoC} device.
This platform hosts a \ac{GPP} and an \ac{FPGA}.
This gives the flexibility to design and develop both hardware, software, and hardware/software co-design paradigms.
For this prototype, we selected the \textit{NaCl core} \cite{Hutter2015} to implement the scalar multiplication, which is one of the key operations in the \ac{DHKE} protocol.

\textit{NaCl Crypto\_box in Hardware} - which we will refer to as ``NaCl core" or simply ``NaCl"- is an example of low-resource hardware implementation of the widely known \textit{crypto\_box} function of the `Networking and Cryptography library' (NaCl) \cite{Hutter2015}.
The NaCl core is in the public domain, making it worthwhile to use.
NaCl uses Curve25519 elliptic curve, which is supported by the popular OpenSSL library and is included in the TLS 1.3 \cite{tls}.
This is the only low-resource hardware implementation of Curve25519 to our knowledge.
The NaCl core supports the X25519 Diffie-Hellman key exchange using Curve25519, the Salsa20 stream cipher, and the Poly1305 message authenticator \cite{Hutter2015}.
The NaCl core is implemented as an Application Specific Instruction Processor (ASIP),
with a silicon area utilization of 14.6k gate equivalent.
It consumes less than 40uW of power consumption at a 1MHz frequency for a 130nm low-leakage CMOS process technology  \cite{Hutter2015}.
There are several reasons why this particular core is chosen.
Firstly, it is a \textit{technology independent} hardware implementation targeting highly \textit{resource-constrained} devices i.e., \textit{optimized for area}.
Secondly, the VHDL code of the core is in the \textit{public domain} and, therefore, freely available for the public.
This allows us to modify it to fit our needs.
Moreover, by using the NaCl core, we build on top of previous academic work and reduce development time.

The performance of the NaCl core mainly depends on the configuration of the multiplier.
The fastest two-cycle version of the core utilizes $2754$ LUT Slices on a Xilinx Artix\textsuperscript{\textregistered}-7 FPGA and takes approximately $830882$ cycles for scalar multiplication.
However, in order to have the smallest possible area utilization, we configure the NaCl core to use a 16 cycle multiplier at the cost of time \cite{Hutter2015}.
Furthermore, the original core contains other functionalities such as the XSalsa20 and Poly1308 code \cite{Hutter2015}, compiled and stored in the ROM program that we do not need for this work.
By reducing the program to its minimum, we further reduced the ROM size by approximately a factor of two.
By doing all this, the final NaCl configuration takes $3.475.123$ cycles and has the lowest area utilization of $946$ LUT Slices on our prototyping platform.

The Xilinx Zynq APSoC platform tightly couples a processor together with the fabric, and the communication is possible via the AXI-peripheral. 
In order to integrate the NaCl core into our prototype setup, we must first wrap the NaCl core into an AXI-peripheral.
Secondly, create \textit{hardware interface drivers} to abstract the hardware and expose only the high-level operations to the programmer.
\subsection{Validation}
\begin{figure}[t]
    \centering
\includegraphics[width=1\textwidth]{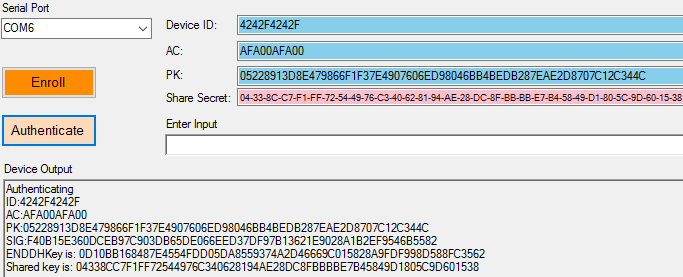}
    \caption{Server GUI; used to \textit{enroll}, \textit{authenticate} and provide the output of the device.}
    \label{fig:gui}
\end{figure}

The essential components needed for all the protocol variants are the same. Therefore, to validate the protocols, we choose to emulate Protocol D on the platform described above, due to its simplicity. We used a server to perform the key-agreement authentication protocol without the device. 
To simplify the verification process, a GUI has been developed that executes the protocol steps. A screenshot of this GUI is provided in Figure~\ref{fig:gui}.
\setlength{\belowcaptionskip}{-10pt}
At the end of the emulation, we verified that both parties (i.e. device and server) derived the same key.
Furthermore, in order to emulate malicious behavior in our prototyping setup, we replaced the authentic SRAM by one that has not been enrolled.
Our experiment showed that we could identify malicious devices and deny their access.


\subsection{Discussion}



In this section, we discuss how this solution satisfies the requirements, as well as its limitations.

\textbf{Mutual authentication key-agreement protocol -}
One of the requirements was to enable secure communication between resource-constrained and unconstrained IoT devices and provide mutual authentication.
Moreover, by using PUF derived keys on the IoT devices, we authenticate silicon.
Hence, an \acf{ECDH} based protocol was designed using the PUF-derived key.
The protocol achieves mutual authentication.
Furthermore, by using a cloud infrastructure, we achieve certificate management that can be used to blacklist devices easily.
Alternatively, based on a particular need, one can choose from three additional protocol variants.

\textbf{Fast time to market -}
We present a generic hardware architecture template that can be used in conjunction with the proposed protocols.
The critical component in the entire system is scalar multiplication.
We show that an off-the-shelf core such as NaCl can be quickly used for such purposes.
Furthermore, this has the potential to be a viable option for a low to medium production, and a fast time to market, which is crucial in the IoT market.
Alternatively, if a higher production volume is required, we might need to design an ASIC.
Although the non-recurring costs are known to be high for such design, the resulting per-unit price can be substantially minimized this way.
Furthermore, ASIC design can be optimized for an area, resulting in the smallest form factor.

\textbf{Minimizing silicon footprint -}
To minimize the footprint, we use \ac{ECC}. 
Due to its shorter operands, area utilization is significantly reduced when compared to other public-key crypto cores, while achieving the same level of security.
Therefore, \ac{ECC} has been gaining popularity in the community as a suitable candidate for constrained devices.
Furthermore, to minimize the silicon footprint of our design, we choose to use SRAM-PUF. SRAM is already present in most systems, and SRAM can be made from standards components.
Furthermore, SRAM-PUF technology itself is gaining popularity, and its widespread adoption is imminent. 

\textbf{Validation -}
In this work, we introduced a systematic approach to building a prototype for the validation of our proposed protocol.
The prototype allowed us to verify, evaluate, and analyze the feasibility of such a system.


\section{Conclusion}
\label{sec:conclusion}
In this paper we proposed a mutual authenticating key-agreement protocol that is based on \ac{ECDH} which uses a SRAM-PUF based key. The proposed protocol is designed for IoT devices that work under stringent constraints, i.e., area, cost and power.
We chose \ac{ECDH} since \ac{ECC} is a suitable candidate for constrained devices due to the shorter operand size. 
Also, \ac{SRAM}-\ac{PUF} is chosen to be used due to its expected adoption in IoT.
In our base protocol, communicating parties are enrolled by a \ac{TTP} to achieve mutual authentication with a negligible \ac{NVM} requirement on the device's side.
Furthermore, we use cloud infrastructure to make certificate management possible.
In order to comply with different scenarios, three additional variants of the original protocol are proposed.
We further provide a  comparison of the variants, showing the trade-offs related to security versus implementation requirements.
As future work, we expect to carry out formal security verification of the proposed protocol and its variants.
The designed protocol served as a roadmap in drafting a modular hardware architecture.
This architecture was prototyped, verified, and its feasibility and practicality were demonstrated on a Xilinx Zynq-7000 APSoC device.


\bibliographystyle{IEEEtran}
\bibliography{ref,biblio}

\begin{thebibliography}{10}
\providecommand{\url}[1]{#1}
\csname url@samestyle\endcsname
\providecommand{\newblock}{\relax}
\providecommand{\bibinfo}[2]{#2}
\providecommand{\BIBentrySTDinterwordspacing}{\spaceskip=0pt\relax}
\providecommand{\BIBentryALTinterwordstretchfactor}{4}
\providecommand{\BIBentryALTinterwordspacing}{\spaceskip=\fontdimen2\font plus
\BIBentryALTinterwordstretchfactor\fontdimen3\font minus
  \fontdimen4\font\relax}
\providecommand{\BIBforeignlanguage}[2]{{%
\expandafter\ifx\csname l@#1\endcsname\relax
\typeout{** WARNING: IEEEtran.bst: No hyphenation pattern has been}%
\typeout{** loaded for the language `#1'. Using the pattern for}%
\typeout{** the default language instead.}%
\else
\language=\csname l@#1\endcsname
\fi
#2}}
\providecommand{\BIBdecl}{\relax}
\BIBdecl

\bibitem{Singh:1999:CBE:519693}
S.~Singh, \emph{The Code Book: The Evolution of Secrecy from Mary, Queen of
  Scots, to Quantum Cryptography}, 1st~ed.\hskip 1em plus 0.5em minus
  0.4em\relax New York, NY, USA: Doubleday, 1999.

\bibitem{IoTproblem}
\BIBentryALTinterwordspacing
A.~Gerber, \emph{Top 10 IoT security challenges}, 2017 (accessed December,
  2018). [Online]. Available:
  \url{https://developer.ibm.com/articles/iot-top-10-iot-security-challenges/}
\BIBentrySTDinterwordspacing

\bibitem{Lee2004}
\BIBentryALTinterwordspacing
J.~W. Lee, D.~Lim, B.~Gassend, G.~E. Suh, M.~van Dijk, and S.~Devadas, ``{A
  technique to build a secret key in integrated circuits for identification and
  authentication applications},'' in \emph{2004 Symposium on VLSI Circuits.
  Digest of Technical Papers (IEEE Cat. No.04CH37525)}, 2004, pp. 176--179.
  [Online]. Available: \url{http://ieeexplore.ieee.org/document/1346548/}
\BIBentrySTDinterwordspacing

\bibitem{Suh2007}
G.~E. Suh and S.~Devadas, ``{Physical unclonable functions for device
  authentication and secret key generation},'' in \emph{Proceedings - Design
  Automation Conference}, 2007, pp. 9--14.

\bibitem{Herder2014}
\BIBentryALTinterwordspacing
C.~Herder, M.-D. Yu, F.~Koushanfar, and S.~Devadas, ``{Physical Unclonable
  Functions and Applications: A Tutorial},'' \emph{Proceedings of the IEEE},
  vol. 102, no.~8, pp. 1126--1141, aug 2014. [Online]. Available:
  \url{http://ieeexplore.ieee.org/document/6823677/}
\BIBentrySTDinterwordspacing

\bibitem{Frikken2009}
K.~B. Frikken, M.~Blanton, and M.~J. Atallah, ``{Robust Authentication Using
  Physically Unclonable Functions},'' in \emph{Information Security},
  P.~Samarati, M.~Yung, F.~Martinelli, and C.~A. Ardagna, Eds.\hskip 1em plus
  0.5em minus 0.4em\relax Berlin, Heidelberg: Springer Berlin Heidelberg, 2009,
  pp. 262--277.

\bibitem{rfe1}
A.~Van~Herrewege, S.~Katzenbeisser, R.~Maes, R.~Peeters, A.-R. Sadeghi,
  I.~Verbauwhede, and C.~Wachsmann, ``Reverse fuzzy extractors: Enabling
  lightweight mutual authentication for puf-enabled rfids,'' in \emph{Financial
  Cryptography and Data Security}, A.~D. Keromytis, Ed.\hskip 1em plus 0.5em
  minus 0.4em\relax Berlin, Heidelberg: Springer Berlin Heidelberg, 2012, pp.
  374--389.

\bibitem{Maes2013}
\BIBentryALTinterwordspacing
R.~Maes, \emph{PUF-Based Entity Identification and Authentication}.\hskip 1em
  plus 0.5em minus 0.4em\relax Berlin, Heidelberg: Springer Berlin Heidelberg,
  2013, pp. 117--141. [Online]. Available:
  \url{https://doi.org/10.1007/978-3-642-41395-7_5}
\BIBentrySTDinterwordspacing

\bibitem{AGMSY15}
A.~Aysu, E.~Gulcan, D.~Moriyama, P.~Schaumont, and M.~Yung, ``End-to-end design
  of a puf-based privacy preserving authentication protocol,'' in
  \emph{Cryptographic Hardware and Embedded Systems -- CHES 2015},
  T.~G{\"u}neysu and H.~Handschuh, Eds.\hskip 1em plus 0.5em minus 0.4em\relax
  Berlin, Heidelberg: Springer Berlin Heidelberg, 2015, pp. 556--576.

\bibitem{Barbareschi2018}
\BIBentryALTinterwordspacing
M.~Barbareschi, A.~{De Benedictis}, and N.~Mazzocca, ``{A PUF-based hardware
  mutual authentication protocol},'' \emph{Journal of Parallel and Distributed
  Computing}, vol. 119, pp. 107--120, 2018. [Online]. Available:
  \url{https://doi.org/10.1016/j.jpdc.2018.04.007}
\BIBentrySTDinterwordspacing

\bibitem{Aysu2016}
\BIBentryALTinterwordspacing
A.~Aysu, E.~Gulcan, D.~Moriyama, and P.~Schaumont, ``{Compact and low-power
  ASIP design for lightweight PUF-based authentication protocols},'' \emph{IET
  Information Security}, vol.~10, no.~5, pp. 232--241, 2016. [Online].
  Available:
  \url{http://digital-library.theiet.org/content/journals/10.1049/iet-ifs.2015.0401}
\BIBentrySTDinterwordspacing

\bibitem{Hutter2015}
M.~Hutter, J.~Schilling, P.~Schwabe, and W.~Wieser, \emph{NaCl's Crypto{\_}box
  in Hardware}.\hskip 1em plus 0.5em minus 0.4em\relax Berlin, Heidelberg:
  Springer Berlin Heidelberg, 2015, pp. 81--101.

\bibitem{Bernstein2006}
D.~J. Bernstein, ``{Curve25519: New Diffie-Hellman Speed Records},'' in
  \emph{Public Key Cryptography - PKC 2006}, M.~Yung, Y.~Dodis, A.~Kiayias, and
  T.~Malkin, Eds.\hskip 1em plus 0.5em minus 0.4em\relax Berlin, Heidelberg:
  Springer Berlin Heidelberg, 2006, pp. 207--228.

\bibitem{Ravikanth:2001:POF:935173}
P.~S. Ravikanth, ``Physical one-way functions,'' Ph.D. dissertation, Cambridge,
  MA, USA, 2001, aAI0803255.

\bibitem{DBLP:books/sp/Maes13}
\BIBentryALTinterwordspacing
R.~Maes, \emph{Physically Unclonable Functions - Constructions, Properties and
  Applications}.\hskip 1em plus 0.5em minus 0.4em\relax Springer, 2013.
  [Online]. Available: \url{http://dx.doi.org/10.1007/978-3-642-41395-7}
\BIBentrySTDinterwordspacing

\bibitem{maes2010physically}
R.~Maes and I.~Verbauwhede, ``Physically unclonable functions: A study on the
  state of the art and future research directions,'' in \emph{Towards
  Hardware-Intrinsic Security}.\hskip 1em plus 0.5em minus 0.4em\relax
  Springer, 2010, pp. 3--37.

\bibitem{cryptoeprint:2015:583}
R.~Maes, V.~van~der Leest, E.~van~der Sluis, and F.~Willems, ``Secure key
  generation from biased pufs,'' Cryptology ePrint Archive, Report 2015/583,
  2015, \url{http://eprint.iacr.org/}.

\bibitem{6513684}
V.~van~der Leest and P.~Tuyls, ``Anti-counterfeiting with hardware intrinsic
  security,'' in \emph{Design, Automation Test in Europe Conference Exhibition
  (DATE), 2013}, March 2013, pp. 1137--1142.

\bibitem{Maes2010}
\BIBentryALTinterwordspacing
R.~Maes and I.~Verbauwhede, \emph{Physically Unclonable Functions: A Study on
  the State of the Art and Future Research Directions}.\hskip 1em plus 0.5em
  minus 0.4em\relax Berlin, Heidelberg: Springer Berlin Heidelberg, 2010, pp.
  3--37. [Online]. Available: \url{https://doi.org/10.1007/978-3-642-14452-3_1}
\BIBentrySTDinterwordspacing

\bibitem{modelmafalda}
A.~Cortez, A.~Dargar, G.~Schrijen, and S.~Hamdioui, ``Modeling sram start-up
  behavior for physical unclonable functions,'' in \emph{Proc. IEEE
  International Symposium on Defect and Fault Tolerance in VLSI and
  Nanotechnology Systems}, Austin, USA, October 2012, pp. 1--6.

\bibitem{kumar2006elliptic}
S.~S. Kumar, ``Elliptic curve cryptography for constrained devices,'' Ph.D.
  dissertation, Ruhr University Bochum, 2006.

\bibitem{Paar:2009:UCT:1721909}
C.~Paar and J.~Pelzl, \emph{Understanding Cryptography: A Textbook for Students
  and Practitioners}, 1st~ed.\hskip 1em plus 0.5em minus 0.4em\relax Springer
  Publishing Company, Incorporated, 2009.

\bibitem{Hugo2010}
H.~Krawczyk, ``Cryptographic extraction and key derivation: The hkdf scheme,''
  in \emph{Advances in Cryptology -- CRYPTO 2010}, T.~Rabin, Ed.\hskip 1em plus
  0.5em minus 0.4em\relax Berlin, Heidelberg: Springer Berlin Heidelberg, 2010,
  pp. 631--648.

\bibitem{Boyen:2004:RCF:1030083.1030096}
\BIBentryALTinterwordspacing
X.~Boyen, ``Reusable cryptographic fuzzy extractors,'' in \emph{Proceedings of
  the 11th ACM Conference on Computer and Communications Security}, ser. CCS
  '04.\hskip 1em plus 0.5em minus 0.4em\relax New York, NY, USA: ACM, 2004, pp.
  82--91. [Online]. Available: \url{http://doi.acm.org/10.1145/1030083.1030096}
\BIBentrySTDinterwordspacing

\bibitem{Dodis2004}
Y.~Dodis, L.~Reyzin, and A.~Smith, \emph{Fuzzy Extractors: How to Generate
  Strong Keys from Biometrics and Other Noisy Data}.\hskip 1em plus 0.5em minus
  0.4em\relax Berlin, Heidelberg: Springer Berlin Heidelberg, 2004, pp.
  523--540.

\bibitem{Linnartz2003}
J.-P. Linnartz and P.~Tuyls, \emph{New Shielding Functions to Enhance Privacy
  and Prevent Misuse of Biometric Templates}.\hskip 1em plus 0.5em minus
  0.4em\relax Berlin, Heidelberg: Springer Berlin Heidelberg, 2003, pp.
  393--402.

\bibitem{tls}
``{The Transport Layer Security (TLS) Protocol Version 1.3},'' Internet
  Engineering Task Force, Standard, accessed on 05-12.2018.

\end{thebibliography}

\vspace{2cm}

\end{document}